\documentclass[a4paper,11pt]{article}
\pdfoutput=1 
\usepackage{jheppub}  

\usepackage[T1]{fontenc}
\usepackage[latin9]{inputenc}
\usepackage[latin9]{inputenc}
\usepackage{amsmath}
\usepackage{amssymb}
\usepackage{esint}
\usepackage{graphicx}
\usepackage{ dsfont }
\usepackage{accents}
\newlength{\dhatheight}

\newcommand{\doublehat}[1]{%
	\settoheight{\dhatheight}{\ensuremath{\hat{#1}}}%
	\addtolength{\dhatheight}{-0.35ex}%
	\hat{\vphantom{\rule{1pt}{\dhatheight}}%
		\smash{\hat{#1}}}}

\newcommand{\doublehatexp}[1]{%
	\settoheight{\dhatheight}{\ensuremath{\hat{#1}}}%
	\addtolength{\dhatheight}{-0.77ex}%
	\hat{\vphantom{\rule{1pt}{\dhatheight}}%
		\smash{\hat{#1}}}}



\title{\boldmath Conformal Bootstrap near the edge}

\author[a]{Ant\'{o}nio Antunes}

\affiliation[a]{Centro de F\'\i sica do Porto,
	Departamento de F\'\i sica e Astronomia\\
	Faculdade de Ci\^encias da Universidade do Porto\\
	Rua do Campo Alegre 687,
	4169--007 Porto, Portugal}

\emailAdd{alantunes@fc.up.pt}

\abstract{We propose a bootstrap program for CFTs near intersecting boundaries which form a co-dimension 2 edge. We describe the kinematical setup and show that bulk 1-pt functions and bulk-edge 2-pt functions depend on a non-trivial cross-ratio and on the angle between the boundaries. Using the boundary OPE (BOE) with respect to each boundary, we derive two independent conformal block expansions for these correlators. The matching of the two BOE expansions leads to a crossing equation. We analytically solve this equation in several simple cases, notably for a free bulk field, where we recover Feynman-diagrammatic results by Cardy.}

\begin{document} 
	\setlength{\abovedisplayskip}{5pt}
	\setlength{\belowdisplayskip}{5pt}
	\maketitle
	\flushbottom
\section{Introduction}
The modern conformal bootstrap program has led to remarkable progress in the understanding of conformal field theories in space-time dimensions greater than two \cite{Rattazzi:2008pe,Poland:2018epd}. The simple but powerful idea of the method is to impose conformal invariance, unitarity and Bose/Fermi symmetry on 4-pt correlation functions, leading to non-trivial, theory independent constraints on the CFT data and even allowing for the solution of specific models after providing additional assumptions. This has been particularly relevant in the context of critical phenomena, since the unprecedentedly accurate predictions on the scaling dimensions of local operators and OPE coefficients, the so-called CFT data, translate into benchmark values for critical exponents and structure constants.

From the point of view of statistical mechanics, and second order phase transitions, it is very natural to study more general setups where part of the conformal symmetry is broken. For example, an experimentalist might want to measure his critical sample near its surface. In particular, a system can exhibit different types of criticality regarding its surface and bulk degrees freedom, leading for example to different critical exponents. An interesting example of this is the phase diagram of the Ising model with a surface interaction. The extension of conformal field theory to this setup is known as boundary conformal field theory (BCFT) \cite{Cardy:1984bb,Cardy:1989ir,Cardy:1991tv,Cardy:2004hm,Diehl:1981zz}. Aside from containing the same local degrees of freedom and observables of the bulk theory, BCFT additionally contains local operators living on the boundary. This means that the CFT data further includes the scaling dimensions of boundary operators and the coefficients of the expansion of bulk degrees of freedom in terms of their boundary counterpart (BOE) \cite{McAvity:1993ue,McAvity:1995zd}. Remarkably, the consistency of the bulk operator product expansion with the boundary operator expansion leads to a crossing equation which imposes powerful non-perturbative constraint on the bulk and boundary CFT data, extending the applicability of the conformal bootstrap philosophy \cite{Liendo:2012hy,Gliozzi:2015qsa,Mazac:2018biw,Kaviraj:2018tfd,Bissi:2018mcq,Dey:2020lwp,Dey:2020jlc,Shpot:2019iwk,Bianchi:2019sxz}. 

The extension of this program to defects of arbitrary co-dimension, known as defect CFT, has also had similar success \cite{Billo:2016cpy,Gadde:2016fbj,Gaiotto:2013nva,Liendo:2016ymz,Liendo:2018ukf,Billo:2013jda,Fukuda:2017cup}. In the case of co-dimension higher than one, the transverse rotation symmetry of the defect plays an interesting role as it becomes a global internal symmetry from the point of view of the defect local operators, organizing them in representations of the transverse rotation group \cite{Lemos:2017vnx,Liendo:2019jpu,Lauria:2018klo,Bianchi:2018zpb}.

We also note that other mild modifications of conformal symmetry have proved to be just as powerful in teaching us about the rich properties of CFTs. Notably, the study of CFT at finite temperature, which is tantamount to probing the theory in the manifold $\mathbb{R}^{d-1}\times S^1$, along with the periodicity condition for correlators in this geometry (i.e. the KMS condition), leads to a set of bootstrap equations constraining the thermal data \cite{Iliesiu:2018fao,Iliesiu:2018zlz}. We emphasize that this setup introduces an explicit dimensionful scale to the system, whose effects are somewhat tamed by the periodicity. Additionally, CFTs in the background of a real projective space have also been studied, leading to results which are quite similar in nature to the BCFT case \cite{Giombi:2020xah,Nakayama:2016cim}.

This finally leads us to the case at hand, a conformal field theory probed by two intersecting boundaries. Parallel boundaries, or defects, lead to the introduction of an explicit length scale destroying all hopes to take advantage of the full power of conformal symmetry \cite{Soderberg:2021kne}. Intersecting boundaries however, lead to a type of deformation of conformal symmetry qualitatively different from all the examples discussed above. On the one hand, it does not introduce any length scales, making it qualitatively different from thermal CFT. On the other hand it introduces a dimensionless parameter, $\theta$ the angle between the two boundaries, as opposed to BCFT or defect CFT which are sharp, rigid deformations of homogenous CFT. We remark that even thermal CFT is not a one parameter deformation, since the deformation parameter is dimensionful, meaning all non-zero values of temperature are equivalent in a CFT.
We have arrived then at the two main motivations for studying CFT in a wedge:
\begin{itemize}
	\item[$(i)$] Experimental and computational critical systems have boundaries and edges.
	\item[$(ii)$] Introducing a wedge of angle $\theta$ is a one-parameter deformation of a CFT (albeit disconnected from the homogeneous case).
\end{itemize}
There is also an important historical motivation. In the 1980's many critical systems were studied in a wedge configuration. Notably, Cardy attacked this problem for $O(N)$ models in the $4-\epsilon$ expansion \cite{Cardy_1983}, which lead to other developments, including in 2 and 3 dimensional systems \cite{Guttmann_1984,Barber1984,Cardy_1984,Kaiser1989,Pleimling_1998}. The results by Cardy will serve as a guiding principle in many points of this work.

With this incentive, we now propose to apply the conformal bootstrap approach one more time. We introduce edge scaling dimensions, and boundary to edge expansion coefficients. Imposing compatibility of the boundary expansion on the two boundaries will lead to consistency equations relating the data of the bulk, the two boundaries and the edge. This leads to a rich setup, which contains one bulk theory with a reduced conformal symmetry, two boundary theories, themselves BCFTs, since the edge plays the role of the boundary of a boundary, and an edge theory, with the full conformal symmetry for a $d-2$ dimensional theory.

The paper is structured as follows. We begin in section \ref{sec:kinematics} by carefully describing the setup and analyzing the relevant kinematics. In section \ref{sec:BOEexpansion} we take advantage of the boundary operator expansion, developing a conformal block expansion for the bulk one point functions. Imposing consistency of the two boundary expansions leads to a crossing equation, analogous to the ones in BCFT or homogeneous CFT. In section \ref{sec:1ptcrossing} we analyze the properties of the crossing equation and solve them in simple cases, notably in the case where the bulk field has the dimension of a free scalar field. In section \ref{sec:bulkedgecrossing}, we extend the previous program to the case where one considers a bulk-edge two point function, making a connection to the results by Cardy. We conclude and discuss future avenues in section \ref{sec:Conclusions}.

\section{Kinematical Setup}
\label{sec:kinematics}
\begin{figure}[h]
 	\centering
 	\includegraphics[width=0.7\linewidth]{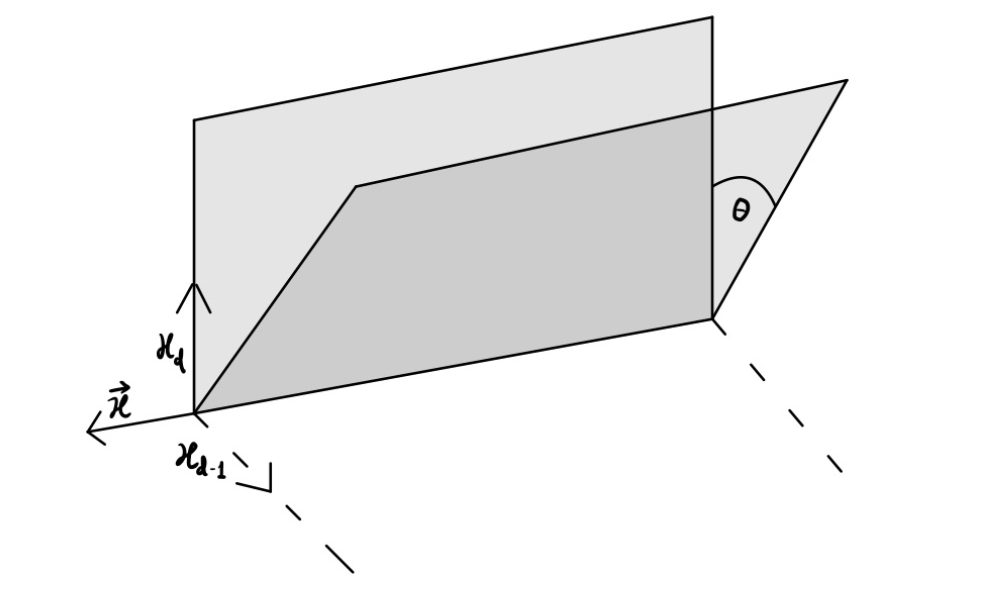}
 	\caption{Setup for CFT near two intersecting boundaries forming an angle $\theta$.}
 	\label{fig:twoboundaries}
 \end{figure}

We consider a $d$-dimensional CFT near two intersecting boundaries, which form an edge of co-dimension 2. We take the normal vectors of the boundaries to live in the $x_{d-1},x_d$ plane, and let the surfaces have an angle $\theta$, with one of the boundaries, taken conventionally at $x_{d-1}=0$. Note that in the limit $\theta \to \pi$ we recover the usual BCFT configuration. We label the directions along the co-dimension 2 edge by $\vec{x}$. We present the setup in figure \ref{fig:twoboundaries}.

 Let us now analyse the symmetry of this system. First recall that a usual bulk CFT possesses SO($d+1,1$) symmetry, generated by $d$ translations, $d$ special conformal transformations, 1 dilation and $d(d-1)/2$ rotations. This adds up to $(d+2)(d+1)/2$ generators. By introducing one boundary, we break translation symmetry and the associated SCT of the direction normal to the boundary. Furthermore we can no longer perform rotations that change the normal vector, so we have $d-1$ fewer rotations allowed. This gives a theory with $d-1$ translations $d-1$ SCTs, 1 dilation and $(d-1)(d-2)/2$ rotations, which shows that BCFTs have SO($d,1$) symmetry, as is well known. Importantly the boundary is scale invariant, because $x_d=0$ is a scale invariant condition, and the remaining SCTs are easily shown to persist, since the system maintains inversion symmetry \cite{McAvity:1995zd,Cardy:1984bb}.
 
 Now, the introduction of a second, intersecting and non-coincident boundary breaks an additional translation, the associated SCT, and $d-2$ rotations, since rotations involving only the $x_{d-1}$ and $x_d$ coordinates were already broken by the "first" boundary. Clearly scale invariance and inversion symmetry remain, since the BCFT derivation holds for both boundaries simultaneously.
 We are left then with SO($d-1,1$) symmetry, which means the system still has some leftover conformal invariance for $d>2$. In particular, the theory on the edge has the full symmetry of a CFT in the appropriate $d-2$ dimensions. The case $d=2$ leaves only scale invariance, and we therefore assume $d>2$ from now on. We also emphasize that $\theta$ is an external parameter of our setup that we can tune as we please. This means that the edge CFT data generically depends on $\theta$.
 \subsection{Embedding Formalism and wedge correlation functions}
 We now adapt the embedding space formalism \cite{Costa:2011mg} to this setup. This will clarify the SO($d-1,1$) invariance and allow us to trivially write down the general form of bulk 1-pt functions. Consider the embedding formalism for SO($d+1,1$) acting linearly on the coordinates of $\mathbb{R}^{d+1,1}$ and consider the projective null cone
 \begin{equation}
 P^{A}=(P^+,P^-,P^1,\dots,P^d) \,,\quad P^AP_A=0 \,,\quad P^A \sim \lambda P^A \,;\quad \lambda>0\,,
 \end{equation}
 Physical space is obtained by $x^\mu= P^\mu/P^+$
The presence of a boundary at $x_{d-1}=0$ is implemented by introducing a vector \cite{Liendo:2012hy}
\begin{equation}
V_1^A= (0,\dots,0,1,0)\,,
\end{equation}
which selects a special direction that must be preserved by conformal transformations.
 The other boundary is implemented by introducing a second vector\footnote{One might want to introduce a vector $V_\theta^A= (0,\dots,0,-\cos \theta,\sin \theta)$, normal to the tilted boundary. However, since our observables will anyway explicitly depend on $\theta$, we can just replace it by $V_2$. Clearly, transformations that leave $V_1$ and $V_2$ invariant also leave $V_1$ and $V_\theta$ invariant and vice-versa.}
 \begin{equation}
 V_2^A=(0,\dots,0,0,1)\,.
 \end{equation}
 It is now clear that rotations that don't touch the last two coordinates leave the system invariant, making manifest the SO($d-1,1$) symmetry. Let us consider then a 1-pt function of a scalar operator
 \begin{equation}
 	\langle \mathcal{O}(\vec{x},x_{d-1},x_d)\rangle\,.
 \end{equation}
 In embedding space we promote the fields to be homogeneous functions of $P$, with
 \begin{equation}
 	\mathcal{O}(\lambda P)= \lambda^{-\Delta} \mathcal{O}(P)\,.
 \end{equation} 
 This means we must construct a homogeneous function of degree $-\Delta$ in $P$ using $V_1,V_2$ and $P$. This fixes the form of the correlator to be
 \begin{equation}
 	\langle\mathcal{O}(P)\rangle= \frac{f\left( \frac{P\cdot V_1}{P\cdot V_2},\theta\right) }{(2P\cdot V_1)^\Delta}\,,
 \end{equation}
 where we conventionally chose the prefactor to be $(2P\cdot V_1)^{-\Delta}$. Other choices, such as $(2P\cdot V_2)^{-\Delta}$ are related by multiplication by a function of the cross ratio $(P\cdot V_1)/(P \cdot V_2)$. Upon projection to physical space we obtain
 \begin{equation}
 		\langle\mathcal{O}(\vec{x},x_{d-1},x_d)\rangle = \frac{f(\eta,\theta)}{(2x_{d-1})^\Delta}\,,
 \end{equation}
 where we introduce the cross ratio $\eta$ defined as
 \begin{equation}
 \eta= \frac{x_{d-1}}{x_d} \equiv \tan \phi \,.
 \end{equation}
This means that a 1-pt function for edge CFT is non-trivial, because of the kinematical angular dependence in $\phi$ and the parametric dependence in $\theta$. The explicit breaking of the transverse rotation symmetry around the edge means that the $d-2$ dimensional theory is qualitatively different from a defect CFT in co-dimension 2 where the defect spectrum organizes in representations of SO(2)\footnote{However, the edge CFT is somewhat reminiscent of the so-called spinning conformal defects \cite{Kobayashi:2018okw}, which are themselves charged under the transverse rotation group. It would be interesting to understand if there is a precise connection between the physics of these two systems.} \cite{Billo:2016cpy,Gadde:2016fbj}. 
A slight generalization of the one point correlator of a bulk field are the bulk-edge two point functions, where we insert an operator $\doublehat{\mathcal{O}}$ (we use two hats for edge operators, one hat for boundary operators and no hats for bulk operators). Symmetry now determines
\begin{equation}
	\langle \mathcal{O}_1(P_1) \doublehat{\mathcal{O}}_2(P_2)\rangle = \frac{f\left( \frac{P\cdot V_1}{P\cdot V_2},\theta\right) }{(-2P_1\cdot P_2)^{\doublehatexp{\Delta}_2}(2P_1\cdot V_1)^{\Delta_1-\doublehatexp{\Delta}_2}}= \frac{f(\phi,\theta)}{r^{2\doublehatexp{\Delta}_2} \left(2x_{1,d-1}\right)^{\Delta_1-\doublehatexp{\Delta}_2} }\,,
\end{equation} 
where $r^2=\vec{x}_{12}^2+x^2_{1,d-1}+x^2_{1,d}$ and $\vec{x}_{12} = \vec{x}_1-\vec{x}_2$.
Note that by using translations we can set $\vec{x}_2=0$. A subsequent special conformal transformations along the edge direction allows us to have $\vec{x}_1=0$ at the cost of changing the perpendicular distance to the edge which can be undone by a scaling transformation. Additionally, note that by setting $\doublehat{\Delta}_2=0$, we can recover the bulk 1-pt function case.
 It will also be convenient to consider the boundary-edge 2-pt function
 \begin{equation}
 \langle \hat{\mathcal{O}}_1(P_1) \doublehat{\mathcal{O}}_2(P_2)\rangle = \frac{\hat{\mu}^1_2(\theta)}{(-4P_1 \cdot P_2)^{\doublehatexp{\Delta}_2}(2P_1 \cdot V_2)^{\hat{\Delta}_1-\doublehatexp{\Delta}_2}} \equiv \frac{\hat{\mu}^1_2(\theta)}{(2\hat{r}^2)^{\doublehatexp{\Delta}_2}(2x_d)^{\hat{\Delta}_1-\doublehatexp{\Delta}_2}}\,,
 \end{equation}
 where $\hat{r}^2= \vec{x}_{12}^2+x_d^2$ since we took the boundary point to be in the boundary at $x_{d-1}=0$. We also chose an unusual factor of 2 in the definition of $\hat{\mu}$ for later convenience. We can also take the edge operator to be the identity by setting $\doublehat{\Delta}_2$ in which case we simply have
 \begin{equation}
 	\langle \hat{\mathcal{O}}_1(P_1)\rangle= \frac{\hat{\mu}^1_{\mathds{1}}(\theta)}{(2P_1 \cdot V_2)^{\hat{\Delta}_1}}= \frac{\hat{\mu}^1_{\mathds{1}}(\theta)}{(2x_d)^{\hat{\Delta}_1}}\,.
 \end{equation}
  A similar formula will hold for the other boundary. The previous formulas highlight the fact that for each $\theta$ the boundary theory is a BCFT, with the edge playing the role of the boundary of the boundary. This is a testament to the richness of the setup, which contains one bulk theory, two boundary theories, themselves BCFTs and an edge theory, with the full conformal symmetry for a $d-2$ dimensional space. We conclude this section with a table describing all 1 and 2 point functions in terms of the CFT data involved and the relevant cross-ratios.
\begin{center}
	\begin{tabular}{|c|c|c|c|c|} 
		\hline
		$\langle\,\,\,\,\rangle$ & $\emptyset$ & Edge $\doublehat{\mathcal{O}}_1(\vec{x}_1)$ & Boundary $\hat{\mathcal{O}}_1(\vec{x}_1,x_{1,d})$ & Bulk $\mathcal{O}_1(\vec{x}_1,x_{1,d-1},x_{1,d})$ \\
		\hline
		$\emptyset$ & $\emptyset$ & $\delta_{1,\mathds{1}}$ & $\frac{\hat{\mu}^1_{\mathds{1}}(\theta)}{(2x_{1,d})^{\hat{\Delta}_1}}$ & $
		 \frac{f(\eta_1,\theta)}{(2x_{1,d})^{\Delta_1}}$
		 \\ 
		\hline
		 $\doublehat{\mathcal{O}}_2$ &  & $\frac{\delta_{\hat{\hat{1}},\hat{\hat{2}}}}{|\vec{x}_{12}|^{2\hat{\hat{\Delta}}_2}}$ &$ \frac{\hat{\mu}^1_2(\theta)}{(2x_{1,d})^{\hat{\Delta}_1-\hat{\hat{\Delta}}_2}(2\hat{r}^2)^{\hat{\hat{\Delta}}_2}}$ & $\frac{g(\eta_1,\theta)}{(2x_{1,d-1})^{\hat{\Delta}_1-\hat{\hat{\Delta}}_2}r^{2\hat{\hat{\Delta}}_2}}$ \\ 
		\hline
		 $\hat{\mathcal{O}}_2$ & &  & $\frac{f(\zeta_{12},\theta)}{(2x_{1,d})^{\hat{\Delta}_1}(2x_{2,d})^{\hat{\Delta}_2}}$ & $\frac{f(\zeta_{12},\eta_1,\theta)}{(2x_{1,d-1})^{\Delta_1}(2x_{2,d})^{\hat{\Delta}_2}} $ \\ 
		\hline
		 $\mathcal{O}_2$ & & & & $\frac{f(\zeta_{12},\eta_1,\eta_2,\theta)}{(2x_{1,d-1})^{\Delta_1}(2x_{1,d-1})^{\Delta_2}} $ \\ 
		\hline
	\end{tabular}
\end{center}
Here, we defined the cross-ratios $\eta_1 = \frac{P_1\cdot V_1}{P_1 \cdot V_2}\,,\,\eta_2 = \frac{P_2 \cdot V_1}{P_2 \cdot V_2}$ and $\zeta_{12}= \frac{-2P_1\cdot P_2}{(P_1\cdot V_2)(P_2\cdot V_2)}$. We remark that the bulk-boundary and bulk-bulk correlation functions are interesting observables, possessing 2 and 3 cross-ratios respectively, but we will only study the bulk 1-pt function and the bulk-edge 2-pt function, which are the simplest non-trivial correlators. Additionally, there are also boundary-boundary correlators, which, if the operators are on the same boundary, reduce to the usual 2-pt functions in BCFT. However, when the operators are on different boundaries, this is a new observable, which should be closely related to the ones we will study in this work\footnote{In particular, using the BOE expansion for one of the operators should lead to a block expansion similar to the ones we will study below, but will generically contain contributions from an infinite number of edge operators.}.
We finally note that the choice of vectors $V_i$ in $\zeta_{12}$ should be adapted according to the boundary at which the boundary operator (if any) is localized. 
\section{Boundary OPE, block expansions and crossing equation}
\label{sec:BOEexpansion}
 With the kinematics in place, we can now use the usual arguments of OPE expansions to derive general properties of the bulk 1-pt function. We will make crucial use of the boundary operator expansion (BOE) with respect to each boundary. The requirement that the two expansions match will lead us to a crossing equation. 
 \subsection{Boundary OPE}
In BCFT one has access to the bulk OPE since this is a local procedure which is insensitive to the existence of the boundary, as long as the two bulk operators involved are closer to themselves than to any other operator, including boundary operators. Additionally one is able to expand bulk operators in terms of boundary operators, using the distance to the boundary as an expansion parameter. This is known as the boundary operator expansion or BOE \cite{McAvity:1995zd}. To perform the expansion in the transverse distance to the boundary one needs to find a boundary hemisphere that contains only the bulk operator. This is the analogue of the bulk spheres that separate two bulk operators using radial quantization. In particular we note that the BOE stops converging if there is a boundary operator inserted "directly below" the bulk operator. Kinematics dictate that only boundary scalars can be exchanged in the BOE \cite{McAvity:1995zd,Liendo:2012hy}. In the case of BCFT, with a boundary at $x_{d-1}=0$ and $d-1$ transverse directions labeled by $\vec{x}$, the BOE has the general structure
\begin{equation}
	\mathcal{O}(\vec{x},x_{d-1})= \frac{a_\mathcal{O}}{(2 x_{d-1})^\Delta} + \sum_{l} \frac{\mu^{\mathcal{O}}_l}{(2x_{d-1})^{\Delta-\hat{\Delta}_l}} D[x_{d-1},\partial_{\vec{x}}] \hat{\mathcal{O}}_l(\vec{x})\,,
\end{equation}
where $D$ is a homogeneous differential operator and $a_O= \mu^{\mathcal{O}}_{\mathds{1}}$ is the 1-pt function coefficient, or equivalently the bulk to boundary identity OPE coefficient. Additionally $\mu^ {\mathcal{O}}_l$ are the general bulk-boundary OPE coefficients.
 \begin{figure}
 	\centering
 	\includegraphics[width=0.5\linewidth]{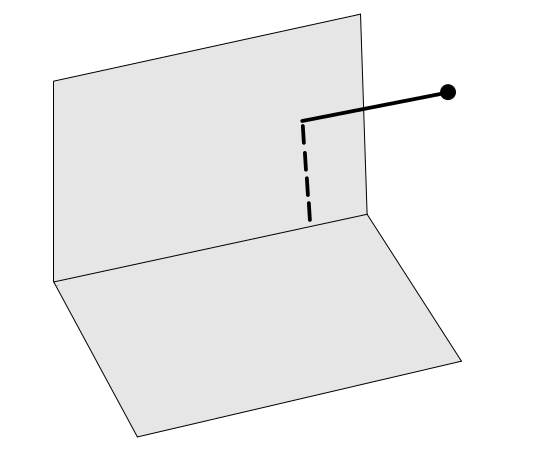}
 	\caption{Diagrammatic representation of the wall channel expansion. The thick line represents the bulk to boundary expansion and the dashed line represents the one point function on the boundary.}
 	\label{fig:wallchann}
 \end{figure}

We can now apply the BOE in our wedge setup.
 Within the region of convergence, which we will discuss below, we can consider a boundary hemisphere, say with respect to the boundary at $x_{d-1}=0$, expanding the bulk operator in a basis of local operators of this boundary. This is essentially a local procedure with respect to the boundary, which is available in spite of the existence of the edge. We call the expansion with respect to the boundary at $x_{d-1}=0$ the wall channel and we represent it in figure \ref{fig:wallchann}. The wall channel BOE simply reads
 \begin{equation}
	\mathcal{O}(\vec{x},x_{d-1},x_d)= \frac{a_\mathcal{O}}{(2 x_{d-1})^\Delta} + \sum_{l} \frac{\mu_l}{(2x_{d-1})^{\Delta-\hat{\Delta}_l}} D[x_{d-1},\partial_{\vec{x}},\partial_{x_d}] \hat{\mathcal{O}}_l(\vec{x},x_d)\,,
 \end{equation}
 where we emphasized the special role that $x_d$ will play, even though it locally is just another transverse direction from the point of view of the BOE around $x_{d-1}$, along with the remaining $d-2$ directions $\vec{x}$.
 Now, we take into account the global features. Since the boundary operators are themselves in a BCFT, where the boundary of the boundary is the edge, they have non-vanishing 1-pt functions, leading to:
  \begin{equation}
  \langle\mathcal{O}(\vec{x},x_{d-1},x_d)\rangle_\theta= \frac{a_\mathcal{O}(\theta)}{(2 x_{d-1})^\Delta} + \sum_{l} \frac{\mu_l}{(2x_{d-1})^{\Delta-\hat{\Delta}_l}} D[x_{d-1},\partial_{\vec{x}},\partial_{x_d}] \frac{a_{\hat{\mathcal{O}}_l}(\theta)}{(2x_d)^{\hat{\Delta}_l}}\,,
  \end{equation}
   Where we allowed for explicit dependence on the angle between the boundaries, since the 1-pt function can ultimately depend on $\theta$, through the data of the edge theory. Of course when $\theta\to \pi$ we expect to be able to recover the usual BOPE  coefficients.
Clearly, because of its local nature, the differential operator is the same as in usual BCFT. The authors of \cite{McAvity:1995zd} showed that, for a boundary operator of dimension $\hat{\Delta}_l$ the differential operator in BCFT is
\begin{equation}
	D[x_{d-1},\partial_{\vec{x}}] = \sum_{m=0}^{\infty}\frac{1}{m! (\hat{\Delta}_l +\frac{3}{2}-\frac{d}{2})_m} \left( -\frac{1}{4}x_{d-1}^2 \vec{
	\nabla}^2\right) ^m	\,.
\end{equation}  
We simply have to use it with special care to distinguish between the $x_d$ and $\vec{x}$ directions, meaning that our differential operator reads
\begin{equation}
\label{eq:explicitD}
	D[x_{d-1},\partial_{\vec{x}},\partial_{x_d}]=\sum_{m=0}^{\infty}\frac{1}{m! (\hat{\Delta}_l +\frac{3}{2}-\frac{d}{2})_m} \left( -\frac{1}{4}x_{d-1}^2 \left( \vec{
		\nabla}^2+\partial^2_{x_d}\right) \right) ^m	\,.
\end{equation}
\subsection{Conformal blocks in the wall channel}
Armed with the explicit differential operator, we are able to write down a block expansion
\begin{equation}
\langle\mathcal{O}(\vec{x},x_{d-1},x_d)\rangle= \frac{1}{(2 x_{d-1})^\Delta} \left( a_\mathcal{O}(\theta) + \sum_l c_l(\theta) f_{\text{wall}}(\hat{\Delta}_l,\eta)\right)\,,
\end{equation}
where we introduced the coefficients
\begin{equation}
	c_l(\theta) = \mu_l~ a_{\hat{\mathcal{O}}_l}(\theta)\,,
\end{equation}
where $a_{\hat{\mathcal{O}}_l}(\theta)$ is the 1-pt function coefficient of $\hat{\mathcal{O}}$ or equivalently the bulk-to-edge OPE coefficient between the boundary operator $\hat{\mathcal{O}}$ and the edge identity operator (only the CFT data involving edge operators is allowed to depend explicitly on $\theta$).
We also defined the wall-channel conformal block
\begin{equation}
	f_{\text{wall}}(\hat{\Delta}_l,\eta) \equiv (2 x_{d-1})^{\hat{\Delta}_l} D[x_{d-1},\partial_{\vec{x}},\partial_{x_d}] \frac{1}{(2x_d)^{\hat{\Delta}_l}}\,.
\end{equation}
Using the representation (\ref{eq:explicitD}) for the differential operator leads to an infinite sum which we can perform explicitly, obtaining
\begin{equation}
		f_{\text{wall}}(\hat{\Delta}_l,\eta) = \eta^{\hat{\Delta}_l} \, _2F_1\left( \frac{\hat{\Delta}_l}{2},\frac{1+\hat{\Delta}_l}{2};\frac{3}{2}-\frac{d}{2}+\hat{\Delta}_l,-\eta^2\right) \,.
\end{equation}
Note that as $\eta \rightarrow 0$, the block behaves as 
\begin{equation}
		f_{\text{wall}}(\hat{\Delta}_l,\eta) \sim \eta^{\hat{\Delta}_l}\,. 
\end{equation}
This is consistent with the OPE limit $x_{d-1}\to 0$ since
\begin{equation}
	\langle \mathcal{O}(\vec{x},x_{d-1}\to 0,x_d)\rangle \sim \frac{1}{(2x_{d-1})^{\Delta-\hat{\Delta}_l}} \langle \hat{\mathcal{O}}_l(\vec{x},x_d)\rangle= \frac{1}{(2x_{d-1})^{\Delta-\hat{\Delta}_l}(2x_d)^{\hat{\Delta}_l}}= \frac{\eta^{\hat{\Delta}_l}}{(2x_{d-1})^{\Delta}}\,.
\end{equation}
Additionally, we can use the fact that the BOE commutes with the boundary Casimir operator to derive a differential equation for the block. Defining, in embedding space, the hatted coordinates
\begin{equation}
	P^{\hat{A}}=\left(P^+, P^-,P^1,\dots,P^{d-2},P^d \right)\,, 
\end{equation}
We easily write the Casimir operator for SO($d,1$)
\begin{equation}
	\hat{L}^2=-\frac{1}{2} L^{\hat{A}\hat{B}} L_{\hat{A}\hat{B}}\,,\quad L_{\hat{A}\hat{B}}= P_{\hat{A}}\partial_{\hat{B}}-P_{\hat{B}}\partial_{\hat{A}}\,.
\end{equation}
Since the Casimir is the same in a given conformal multiplet, we must have
\begin{equation}
	\hat{L}^2\left( \frac{f_{\hat{\Delta}_l}\left(\frac{P\cdot V_1}{P\cdot V_2},\theta\right)}{(P\cdot V_1)^\Delta}\right)=c_{\hat{\Delta}_l,0} \frac{f_{\hat{\Delta}_l}\left(\frac{P\cdot V_1}{P\cdot V_2},\theta\right)}{(P\cdot V_1)^\Delta} \,,
\end{equation}
where $c_{\hat{\Delta}_l,0}$ is the value of the Casimir for a boundary primary $\hat{\mathcal{O}}_l$
\begin{equation}
	c_{\hat{\Delta}_l,0}=\hat{\Delta}_l(\hat{\Delta}_l-d+1)\,.
\end{equation}
Performing elementary manipulations in embedding space and projecting to the physical coordinate space, we derive an ODE for the block in terms of the cross-ratio $\eta$
\begin{equation}
	\eta^2  \left(\eta ^2+1\right) f_{\hat{\Delta}_l}''(\eta
	)+\eta\left(2 \eta ^2+2-d\right) f_{\hat{\Delta}_l}'(\eta )+\hat{\Delta}_l (d-\hat{\Delta}_l -1) f_{\hat{\Delta}_l}(\eta )=0\,.
\end{equation} 
The solution of this equation with the boundary condition $f_{\hat{\Delta}_l}(\eta )\sim \eta^{\hat{\Delta}_l}$ as $\eta$ goes to zero is precisely the one obtained above by ressuming the BOE.
 \subsection{Ramp channel blocks and crossing equation}
\begin{figure}
 	\centering
 	\includegraphics[width=0.5\linewidth]{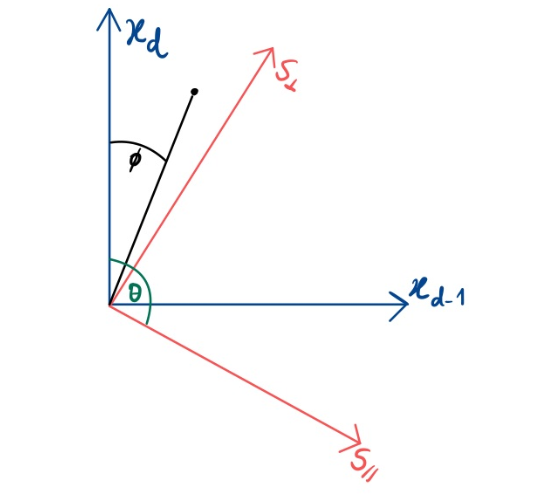}
 	\caption{The two sets of orthogonal coordinates in the wedge setup.}
 	\label{fig:coords}
 \end{figure}

\begin{figure}
 	\centering
 	\includegraphics[width=0.6\linewidth]{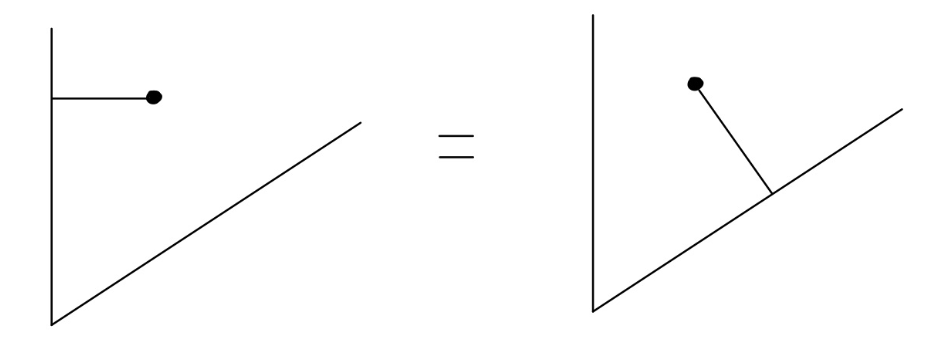}
 	\caption{Diagrammatic representation of the crossing equation for the 1-pt function near an edge. The left hand side crresponds to the wall-channel expansion and the right hand side to the ramp-channel expansion.}
 	\label{fig:crossing}
 \end{figure}

 Having developed the BOE with respect to the boundary at $x_{d-1}$,
 we can now consider the other BOE as the bulk operator approaches the angled boundary. Clearly, if we rotate our axis, this is the same (up to orientation) as the wall channel OPE when we replace $x_{d-1}\to s_\perp$ and $x_d \to s_\parallel$, where $s_\perp$ and $s_\parallel$ are the distances from the insertion point perpendicularly to the angled boundary and the distance along the angled boundary to the edge, respectively. They are given by
\begin{equation}
 	s_\perp=  x_{d}\sin \theta - x_{d-1}\cos \theta  \,,\quad s_\parallel= x_{d}\cos \theta  + x_{d-1}\sin \theta \,,
\end{equation}
we depict the different sets of coordinates in figure \ref{fig:coords}.
It is convenient then to define the cross-ratio with respect to the tilted boundary
\begin{equation}
	\zeta(\theta) \equiv\frac{s_\perp}{s_\parallel}= \frac{\sin \theta-\eta \cos \theta}{\cos \theta+\eta \sin \theta} = \tan(\theta-\phi)\,. 
\end{equation}
Note that $\zeta$ satisfies the expected properties in simple limits:
\begin{equation}
	\zeta(\pi)= -\eta \,,\quad \zeta\left( \frac{\pi}{2}\right)  = \frac{1}{\eta}\,.
\end{equation}
With the appropriate replacements, we can now easily write the ramp-channel conformal block expansion
\begin{equation}
	  \langle\mathcal{O}(\vec{x},x_{d-1},x_d)\rangle= \frac{a'_\mathcal{O}(\theta)}{(2 s_\perp)^{\Delta}} + \sum_{m} \frac{\mu'_m}{(2 s_\perp)^{\Delta-\hat{\Delta}_m}} D[s_\perp,\partial_{\vec{x}},\partial_{s_\parallel}] \frac{a'_{\hat{\mathcal{O}}_m}(\theta)}{(2 s_\parallel)^{\hat{\Delta}_m}}\,,
\end{equation}
leading to
\begin{equation}
\langle\mathcal{O}(\vec{x},x_{d-1},x_d)\rangle= \frac{1}{(2 s_\perp)^{\Delta}} \left( a'_\mathcal{O}(\theta) + \sum_m c'_m(\theta) f_{\text{ramp}}(\hat{\Delta}_m,\eta,\theta)\right)\,,
\end{equation}
with the ramp channel block given by
\begin{equation}
	f_{\text{ramp}}(\hat{\Delta}_m,\eta,\theta)= \zeta^{\hat{\Delta}_m} \, _2F_1\left( \frac{\hat{\Delta}_m}{2},\frac{1+\hat{\Delta}_m}{2};\frac{3}{2}-\frac{d}{2}+\hat{\Delta}_m,-\zeta^2\right)\,, 
\end{equation}
where we suppressed the explicit $\theta$ dependence in $\zeta$. We emphasize that although we expect certain classes of solutions where the spectrum and BOE coefficients on each boundary are the same, a generic solution will have a completely different theory living on each boundary\footnote{Clearly, as $\theta \to \pi$, there should be a solution where the two expansions are identical and additionally one reobtains a purely BCFT result
\begin{equation}
  \langle\mathcal{O}(\vec{x},x_{d-1},x_d)\rangle_{\theta=\pi} \equiv \frac{a_\mathcal{O}}{(2x_{d-1})^\Delta}\,,a_\mathcal{O}(\theta)=a'_\mathcal{O}(\theta)= a_\mathcal{O}+ O(\theta-\pi) \,,\, c_l(\theta)=c'_l(\theta)=0 + O(\theta-\pi)
\end{equation}
}.
With these ingredients, we can write down the crossing equation for general $\theta$
\begin{equation}
	a_\mathcal{O}(\theta) + \sum_l c_l(\theta) f_{\text{wall}}(\hat{\Delta}_l,\eta) = \left( \frac{\eta}{\sin \theta-\eta \cos \theta}\right)^{\Delta} \left(a'_\mathcal{O}(\theta) + \sum_m c'_m(\theta) f_{\text{ramp}}(\hat{\Delta}_m,\eta,\theta) \right)\,. 
\label{eq:gencrossing}	
\end{equation}
This equation is diagrammatically represented in figure \ref{fig:crossing}.
 
 Note that there is an interesting special case when the boundaries are perpendicular, i.e. $\theta=\frac{\pi}{2}$, in this case we use the name floor channel instead of ramp channel, and the equation simplifies to
 \begin{equation}
 		a_\mathcal{O}\left( \frac{\pi}{2}\right)  + \sum_l c_l\left( \frac{\pi}{2}\right)  f_{\text{wall}}(\hat{\Delta}_l,\eta) = \eta
 		^\Delta \left(a'_\mathcal{O}\left( \frac{\pi}{2}\right)  + \sum_m c'_m\left( \frac{\pi}{2}\right)  f_{\text{wall}}\left( \hat{\Delta}_m,\frac{1}{\eta}\right)  \right)\,,
 \end{equation}
 where we used that 
 \begin{equation}
  f_{\text{ramp}}(\hat{\Delta}_m,\eta,\theta=\pi/2)= f_{\text{floor}}(\hat{\Delta}_m,\eta) = f_{\text{wall}}\left( \hat{\Delta}_m,\frac{1}{\eta}\right)\,. 
 \end{equation}
In this case, the blocks on the left/wall channel admit a single power-law expansion around $\eta\to 0$, in even powers of $\eta$, while the block on the right/floor channel admit a similar expansion around $\eta \to \infty$. This is reminiscent of the crossing equation for a 2-pt function in BCFT in terms of the bulk and boundary channels \cite{Liendo:2012hy} and, more generally, of analytic studies of the crossing equation \cite{Komargodski:2012ek,Fitzpatrick:2012yx,Caron-Huot:2017vep}. Note also that the block in the ramp/floor channel, has an interesting small $\eta$ behaviour. Tipically, hypergeometric identities predict two separate power series when the argument of the function is large, but in our case, it turns out that they are integer separated, leading to
\begin{equation}
\label{eq:asymptotic}
	f_{\text{floor}}(\hat{\Delta}_m,\eta) \sim b_0 + b_1 \eta + \dots \,,\quad \eta\to 0\,,
\end{equation}
which is a power series with both even and odd powers of $\eta$. This will play a crucial role when solving the crossing equations below. 
\subsection{Comments on BOE convergence}
In the previous section we assumed that the two boundary expansions had a region of mutual convergence, where the crossing equation is valid. It turns out that this region is somewhat subtle, so we make a few comments on this point before proceeding to analyze solutions of the equations.

\begin{figure}
 	\centering
 	\includegraphics[width=0.9\linewidth]{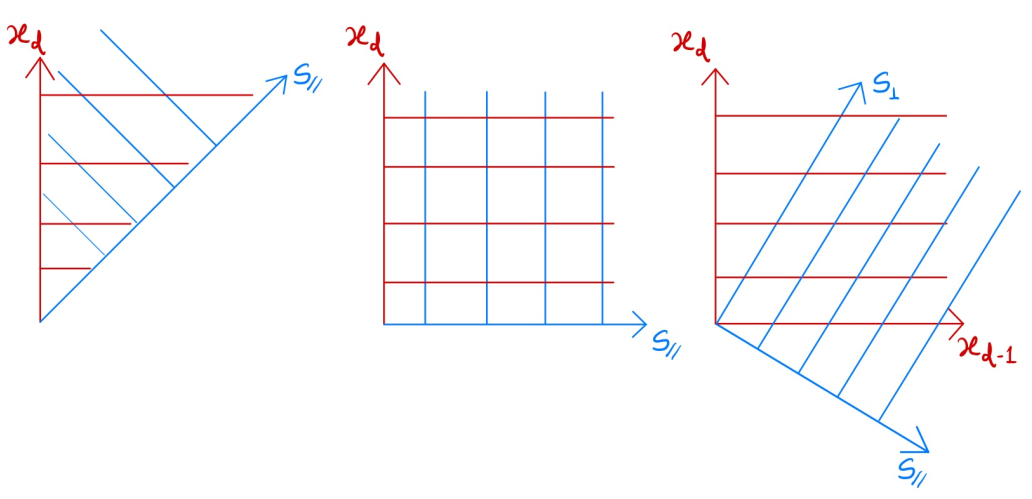}
 	\caption{Regions of convergence for each BOE. Coloured lines represent the region where the associated BOE converges. Lines intersect in the region of mutual convergence}
 	\label{fig:opeconvergence}
 \end{figure}

The crucial aspect to note is that the kinematical region where the two OPEs simultaneously converge depends on theta, and is, in general just a subspace of the full kinematics. For $0<\theta\leq \pi/2$, both BOEs converge for any value of $\phi$ inside the wedge, namely $0<\phi\leq\theta$. However, for an obtuse wedge, only a region centered around $\phi=\theta/2$ ensures convergence in both channels, more precisely $\theta-\pi/2 < \phi< \pi/2$. This can easily be understood by using scale invariance and drawing the usual hemispheres for quantization with respect to each boundaries Hilbert space. By drawing perpendicular lines with respect to each boundary one constructs the tangents of all possible hemispheres centered at the boundary, leading to a sub-wedge where the lines associated to each boundary intersect. This is the region of mutual convergence. We depict the previous procedure in figure \ref{fig:opeconvergence}.
 Therefore, we implicitly work with $\theta\leq \pi/2$, where both BOEs converge inside the full wedge, and analytically continue in $\theta$ when necessary. In particular, the $\theta \to \pi$ limit, which naively recovers the BCFT case, is subtle, since the overlap between the region of convergence of the two expansions vanishes. We also note that $\theta=\pi/2$ is a particularly symmetric case, with the maximum wedge of convergence.
\section{Solving crossing for the 1-pt function}
\label{sec:1ptcrossing}
Having established the validity of the crossing equation (\ref{eq:gencrossing}), we will now attempt to study its possible solutions. 
In general, the bootstrap equation (\ref{eq:gencrossing}) is a non-perturbative constraint on the bulk, boundary and edge CFT data, which contains generically infinitely many unknowns. As in the case of the boundary bootstrap for 2-pt functions, the coefficients of this equation aren't necessarily positive, meaning the standard linear/semi-definite programming approach to the solution of these equations can only be attempted with the assumption of positivity, which is far from general. One could alternatively try to obtain approximate (but uncontrolled) solutions with any sign of the coefficients using Gliozzi's method of determinants. In this work however, we will focus on simple analytically tractable cases and leave the numerical approach for future explorations. 

We will start by looking at a trivial example where only one of the boundaries actually exists.
Subsequently, we will consider some simple but non-trivial regimes which we can study analytically. By taking the bulk field to be a free scalar of dimension $\Delta_d= \frac{d}{2}-1$, we will find that solutions to the crossing equation can contain at most two boundary blocks: $\hat{\Delta}= \frac{d}{2}-1$ and $\hat{\Delta}=\frac{d}{2}$, corresponding to the operators $\hat{\phi}$ and $\partial_\perp\hat{\phi}$, associated to Neumann and Dirichlet boundary conditions. Free boundary conditions correspond to having a single N or D block in each boundary channel. More generally, a combination of these blocks can correspond to non-trivial/interacting boundary conditions for the free bulk field. This was extensively studied in the single boundary case in \cite{Lauria:2020emq,Behan:2020nsf,Prochazka:2019fah}.
\subsection{Warmup: 1-pt function with a single boundary}
\label{sec:oneptsinglebdry}
Let us first consider a one point function where only the boundary at $x_{d-1}$ is present. This case has SO($d,1$) symmetry, and therefore we can expand in our blocks which correspond to a SO($d-1,1$) subgroup. We begin for simplicity by taking $\theta=\pi/2$. The one point function is simply
\begin{equation}
	\langle\mathcal{O}(\vec{x},x_{d-1},x_d)\rangle= \frac{a_{\mathcal{O}}}{(2 x_{d-1})^\Delta}\,,
\end{equation}
which of course means that in the wall channel we only exchange the identity operator with coefficient $a_{\mathcal{O}}$. The crossing equation then reads
\begin{equation}
	\frac{a_\mathcal{O}}{\eta^\Delta} = \sum_n c_n f_{\text{floor}}(\hat{\Delta}_n,\eta)\,.
\end{equation} 
Expanding the equation around $\eta=0$ does not prove useful, since all the blocks behave as a constant. All we learn is that we need infinitely many terms. On the other hand, around $\eta\to\infty$ we have
\begin{equation}
	f_{\text{floor}}(\hat{\Delta}_n,\eta) \sim \eta^{-\hat{\Delta}_n}\left( 1+ O(\eta^{-2})\right)\,,
\end{equation}
 which means that $a'_{\mathcal{O}}=0$ and that the leading operator will be $\hat{\Delta}=\Delta$. This of course creates an infinite tower of terms in $\eta^{-2}$ which we cancel order by order with the addition of operators of dimension $\hat{\Delta}_n=\Delta+2n$. We then find that the coefficients are given by
 \begin{equation}
 	c_n = a_{\mathcal{O}} \frac{4^{-n} (\Delta)_{2n} \Gamma \left(-\frac{d}{2}+n+\Delta
 		+\frac{1}{2}\right)}{n!  \Gamma \left(-\frac{d}{2}+2 n+\Delta
 		+\frac{1}{2}\right)}\,.
 \end{equation}
 The case of arbitrary $\theta$ is similar, except that we must now solve
 \begin{equation}
 	a_{\mathcal{O}} = \left( \frac{\eta}{\sin \theta-\eta \cos \theta}\right)^{\Delta} \sum_m c_m(\theta) f_{\text{ramp}}(\hat{\Delta}_m,\eta,\theta)\,.
 \end{equation}
 Crucially the $\theta$ dependent prefactor leads to odd powers of $\eta^{-1}$, and therefore the expansion contains all operators of the form $\hat{\Delta}_m=\Delta+m$. The coefficients are somewhat more complicated but have the form
 \begin{align}
 	c_m(\theta) &= \sum_{k=0}^{m/2} b_{m,k}\cos(2k\theta) & m \,\text{even}\nonumber\\
 	&= \sum_{k=0}^{(m-1)/2} b'_{m,k}\cos((2k+1)\theta) & m \,\text{odd}\,,
 \end{align}
where $b_{m,k}$ and $b'_{m,k}$ are similar in structure to $c_m$.
This is of course consistent with the case $\theta=\pi/2$, in which case the odd terms are set to zero.
\subsection{Free bulk field with orthogonal boundaries}
Let us now look at a case with a non-trivial boundary spectrum on both boundaries. A simplifying assumption that still leads to interesting physics is to take a free bulk field $\phi$ with dimension $\Delta_d= \frac{d}{2}-1$, in the orthogonal intersection setup. The fact that the bulk field is free does not stop us from having interesting boundary dynamics, as was extensively studied by the authors of \cite{Lauria:2020emq,Behan:2020nsf}. Furthermore, we will see that the edge theory can also present interesting properties.

In this case the crossing equation reads (we suppress the theta dependence of the coefficients since in this section we fix $\theta=\pi/2$):
\begin{align}
	&a_\phi  + \sum_l c_l \eta^{\hat{\Delta}_l} \, F\left( \frac{\hat{\Delta}_l}{2},\frac{1+\hat{\Delta}_l}{2};\frac{3-d+2\hat{\Delta}_l}{2},-\eta^2\right)= \nonumber \\&\eta
	^{\frac{d}{2}-1} \left(a'_\phi + \sum_m \frac{c'_m}{\eta^{\hat{\Delta}_m}}  \, F\left( \frac{\hat{\Delta}_m}{2},\frac{1+\hat{\Delta}_m}{2};\frac{3-d+2\hat{\Delta}_m}{2},-\frac{1}{\eta^{2}}\right) \right)
\end{align}
Now, since the blocks on the right hand side admit a regular series in $\eta$ as $\eta\to0$, we must reproduce a power series of the form $\eta^{\frac{d}{2}-1}\left( k_1 + k_2 \eta + \dots \right)$. This suggests we might be able to reproduce this with a finite number of block on the left hand side. We will generically need two blocks on the left, to account for even and odd powers of $\eta$, and we must set $a_\phi=0$. In particular, we must have $\hat{\Delta}_{l=1}= \frac{d}{2}-1$ to produce the even powers, and $\hat{\Delta}_{l=2}=\frac{d}{2}$ to produce the odd powers. This corresponds to the boundary operators $\hat{\phi}$ and $\partial_\perp \hat{\phi}$, respectively.
Then, for the coefficients of the power series to explicitly match, we must have $a'_\phi=0$ and $\hat{\Delta}_m = \frac{d}{2}-1$ or $\hat{\Delta}_m = \frac{d}{2}$, which can also be seen by expanding at large $\eta$. The most general solution, then, contains $\hat{\phi}$ and $\partial_\perp \hat{\phi}$ on both channels:
\begin{align}
	c_{\hat{\phi}} \eta^{\frac{d}{2}-1} \, F\left( \frac{d}{4}-\frac{1}{2},\frac{d}{4};\frac{1}{2},-\eta^2\right) + c_{\partial_\perp\hat{\phi}} \eta^{\frac{d}{2}} \, F\left( \frac{d}{4},\frac{d}{4}+\frac{1}{2};\frac{3}{2},-\eta^2\right)=\nonumber\\	
c'_{\hat{\phi}} F\left( \frac{d}{4}-\frac{1}{2},\frac{d}{4};\frac{1}{2},-\frac{1}{\eta^2}\right) +\frac{c'_{\partial_\perp\hat{\phi}}}{\eta} F\left( \frac{d}{4},\frac{d}{4}+\frac{1}{2};\frac{3}{2},-\frac{1}{\eta^2}\right)
	\label{eq:freecrossingd}
\end{align}
For these values of the boundary dimensions the blocks simplify. We have
\begin{align}
	\eta^{\frac{d}{2}-1} \, F\left( \frac{d}{4}-\frac{1}{2},\frac{d}{4};\frac{1}{2},-\eta^2\right) &= \sin(\phi)^{\Delta_d} \cos\left( \Delta_d \phi\right)\,, \nonumber\\
	\eta^{\frac{d}{2}} \, F\left( \frac{d}{4},\frac{d}{4}+\frac{1}{2};\frac{3}{2},-\eta^2\right)&= \Delta_d ^{-1} \sin(\phi)^{\Delta_d} \sin\left( \Delta_d \phi\right)\,.
\end{align}
 Furthermore, imposing the precise match of coefficients in the small $\eta$ expansion gives that the primed coefficients are fixed in terms of the unprimed ones, but we still have a two parameter family of solutions constructed in terms of $c_{\hat{\phi}}\,,\, c_{\partial_\perp \hat{\phi}}$. The precise relation is
\begin{align}
	c'_{\hat{\phi}} &= \sin(\pi d/4)c_{\hat{\phi}}-\Delta_d^{-1}\cos(\pi d/4)c_{\partial_\perp \hat{\phi}}\,,\nonumber\\c'_{\partial_\perp \hat{\phi}}&= -\Delta_d\cos(\pi d/4) c_{\hat{\phi}} - \sin(\pi d/4) c_{\partial_\perp \hat{\phi}}\,.
\end{align}
This solution can be easily checked to solve crossing for any value of $\eta$. This is simplest to do in the angular variable $\phi$ where crossing is just $\phi \to \pi/2- \phi$. Let us for a moment take space-time dimension to be 4.
In this case, we can solve the equations with a single block on each side, since they are mapped one-to-one
\begin{equation}
	c'_{\hat{\phi}} =c_{\partial_\perp \hat{\phi}} \,,\, c'_{\partial_\perp \hat{\phi}}=c_{\hat{\phi}}\,.
	\label{eq:d=4perpsol}
\end{equation}
It is interesting to notice that the Dirichlet block gets mapped to the Neumann block and vice-versa. In fact, in this case, the crossing equation simply reads
\begin{equation}
	c_{\hat{\phi}} \frac{\eta}{1+\eta^2} + c_{\partial_\perp \hat{\phi}} \frac{\eta^2}{1+\eta^2} = \eta \left(c'_{\hat{\phi}} \frac{\eta}{1+\eta^2} +c'_{\partial_\perp \hat{\phi}} \frac{1}{1+\eta^2}\right) 
\end{equation}
which is trivially solved by eq. (\ref{eq:d=4perpsol}). A general solution can be obtained by taking any linear combination of the two blocks. 
\subsubsection{Comparison to the equation of motion}
Since the bulk field is free, it satisfies the bulk laplace equation, so we can use this to check the previous results.
For a 1-pt function we simply need to solve the differential equation
\begin{equation}
	\square \langle \phi(\vec{x},x_{d-1},x_d)\rangle=0\,.
\end{equation}
Using the kinematic structure of the point function
\begin{equation}
	\langle \phi(\vec{x},x_{d-1},x_d)\rangle = \frac{f(\frac{x_{d-1}}{x_d}=\eta)}{(2 x_{d-1})^{\frac{d}{2}-1}}\,
\end{equation} 
and that when acting on the $\vec{x}$ independent 1-pt function the laplacian simplifies to
\begin{equation}
	\square \approx \frac{\partial^2}{\partial x_{d-1}^2} + \frac{\partial^2}{\partial x_{d}^2}\,,
\end{equation}
we can derive an ordinary differential equation for $f(\eta)$
\begin{equation}
	4 \eta  \left(\left(2 \eta ^2+2-d\right) f'(\eta )+\eta  \left(\eta ^2+1\right) f''(\eta
	)\right)+d(d-2)f(\eta )=0 \,.
\end{equation}
This is a second order differential equation, and it turns out that the two independent solutions can be written as:
\begin{equation}
	f(\eta)= A f_{\text{wall}}\left( \frac{d}{2}-1,\eta\right)  + B f_{\text{wall}}\left( \frac{d}{2},\eta\right) 
\end{equation}
Which is precisely the combination of Neumann and Dirichlet blocks derived from the crossing equation. This is of course consistent with the fact that we have a two-parameter family of solutions to the crossing equation.

Note that solving the differential equation in $d=4$ leads once again to the simple combination
\begin{equation}
	\langle \phi(\vec{x},x_{d-1},x_d) \rangle = \frac{1}{2 x_{d-1}}\left(c_{\hat{\phi}} \frac{\eta}{1+\eta^2} + c_{\partial_\perp \hat{\phi}} \frac{\eta^2}{1+\eta^2} \right) 
\end{equation}
It is natural from the free field point of view to try to impose free boundary conditions (Neumann or Dirichlet) on each boundary separately. This corresponds to having a single block on each channel which is a subclass of the 2 parameter set of solutions of the crossing equation (\ref{eq:freecrossingd})\footnote{Solutions with a linear combination of both blocks can correspond to interacting boundary theories as discussed in  \cite{Lauria:2020emq,Behan:2020nsf}.}. Imposing N/D BCs at each boundary is achieved by the four possible conditions:
\begin{equation}
	\left(\partial_{x_{d-1,d}}\right) \phi(\vec{x},x_{d-1},x_d) |_{x_{d-1,d}=0}\, 
\end{equation}
meaning we can take the derivative with respect to either $x_{d-1}$ or $x_d$ to vanish in the boundary at $x_{d-1}=0$ or $x_d=0$.
Imposing these boundary conditions leads to the following restrictions on the expansion coefficients
\begin{align}
	D \,,\,x_{d-1}=0 \to c_{\hat{\phi}}= 0 \nonumber\\D\,,\, x_{d}=0 \to c_{\partial_\perp \hat{\phi}}= 0 \nonumber\\N\,,\, x_{d-1}=0 \to c_{\partial_\perp \hat{\phi}}= 0 \nonumber\\ N \,,\, x_{d}=0 \to c_{\hat{\phi}}= 0
\end{align}
Meaning that the only possible free boundary conditions are $DN$ and $ND$, which is consistent with the fact that a single neumann block in one channel corresponds to a single Dirichlet block in the other and vice versa. These boundary conditions intuitively correspond to the fact that at the edge $x_{d-1}=x_d=0$, a parallel derivative in one boundary corresponds to the normal derivative in the other.
\subsubsection{Generalization to arbitrary $\theta$}
It is not hard to generalize the previous results to the case of arbitrary intersection angle $\theta$. We simply use that crossing now sends $\phi \to \theta-\phi$ and account for the $\theta$ dependent prefactor present in equation (\ref{eq:gencrossing}). We can once again write down a solution with only Dirichlet and Neumann blocks on both channels, and expand at small $\eta$ to fix the coefficients. We still find, for each theta, a two-parameter family of solutions given by
\begin{align}
c'_{\hat{\phi}} &= \cos(\Delta_d \theta)c_{\hat{\phi}}+\Delta_d^{-1} \sin(\Delta_d\theta)c_{\partial_\perp \hat{\phi}}\,,\nonumber\\c'_{\partial_\perp \hat{\phi}}&= \Delta_d \sin(\Delta_d\theta) c_{\hat{\phi}} - \cos(\Delta_d\theta) c_{\partial_\perp \hat{\phi}}\,.
\end{align}
Once again, using the $\phi$ variable, we can check that the previous relations solve crossing for any value of the cross-ratio. We can of course recover the orthogonal boundary case by setting $\theta=\pi/2$.

Having the extra parameter $\theta$ to play with, we can find other interesting special solutions. For example, it was impossible to  find a Dirichlet-Dirichlet solution in the orthogonal boundaries case. Now we can consistently set $c_{\hat{\phi}}=c'_{\hat{\phi}}=0$, without making the whole solution vanish. To make this happen, we must have some critical angles $\theta_d$ which take the values
\begin{equation}
	\theta_d = \frac{2 \pi}{d-2}\,.
\end{equation} 
For these angles, we can solve the crossing equation with a single $\partial_\perp \hat{\phi}$ block on each side, and the coefficients satisfy
\begin{equation}
	c'_{\partial_\perp \hat{\phi}} = c_{\partial_\perp \hat{\phi}}\,.
\end{equation}
That is, we can set the free Dirichlet-Dirichlet boundary conditions without trivializing the 1-pt function only for certain special angles $\theta_d$. We note that there are no interesting DD one point functions for $d\leq4$ since $\theta_3=2\pi$ and $\theta_4=\pi$.
\section{Bulk-edge 2-pt function}
\label{sec:bulkedgecrossing}
In the previous section we showed that generically, we cannot impose DD boundary conditions in a 1-pt function of a free bulk field. Such boundary conditions are very natural from the Feynman perturbation theory point of view. In fact, Cardy \cite{Cardy_1983} studied the Wilson-Fisher fixed point in the wedge geometry geometry precisely by deriving free theory propagators for the bulk field with Dirichlet-Dirichlet boundary conditions. In particular he derived interesting critical exponents for correlators where one or both of the bulk fields are close to the boundary. This suggests that we can access interesting CFT data and a bigger set of boundary conditions, including the DD case, by considering a slightly more general correlator. We will consider the simplest non-trivial 2-pt function which is the bulk-edge 2-pt function. As discussed in section \ref{sec:kinematics}, this depends again on a single cross-ratio, but crucially introduces an extra parameter, the dimension of the edge operator $\doublehat{\Delta}$, which can be seen as a function of $\theta$. 
\subsection{Block expansion and crossing equation}
With this in mind, we can start from the bulk-edge correlator, use translational inariance to set $\vec{x}_2=0$ and use the BOE in the wall channel to reduce the calculation to an infinite sum of boundary-edge two point functions
\begin{align}
\label{eq:bulkedgefirstexp}
		\langle \mathcal{O}_1(\vec{x},x_{d-1},x_d) \doublehat{\mathcal{O}}_2(0)\rangle&=  \sum_{l} \frac{\mu^1_l}{(2x_{d-1})^{\Delta_1-\hat{\Delta}_l}} D[x_{d-1},\partial_{\vec{x}},\partial_{x_d}] \langle\hat{\mathcal{O}}_l(\vec{x},0,x_d)  \doublehat{\mathcal{O}}_2(0)\rangle\nonumber\\
		&=\sum_{l} \frac{\mu^1_l }{(2x_{d-1})^{\Delta_1-\hat{\Delta}_l}} D[x_{d-1},\partial_{\vec{x}},\partial_{x_d}] \frac{\hat{\mu}^l_2}{(2 x_d)^{\hat{\Delta}_l - \doublehatexp{\Delta}_2}(2\hat{r}^2)^{\doublehatexp{\Delta}_2}}\,.
\end{align}
Note that when the $l= \mathds{1}$ (the boundary identity operator), we are evaluating an edge 1-pt function, which is non-vanishing only for the edge identity operator. Also, we can easily recover the bulk 1-pt expansion when we set $\doublehat{\Delta}_2=0$. As discussed above, we can always do a conformal transformation to set $\vec{x}=0$, simplifying the analysis. However this should only be done after computing the transverse derivatives in the BOE. Proceeding with the calculation leads to a slight modification of the block expansion derived above for the bulk 1-pt function
\begin{equation}
\label{eq:bulkedgeexp}
	\langle \mathcal{O}_1(\vec{0},x_{d-1},x_d) \doublehat{\mathcal{O}}_2(0)\rangle = \frac{1}{(2 x_{d-1})^{\Delta_1-\doublehatexp{\Delta}_2}r^{2\doublehatexp{\Delta}_2}}\sum_l c_l^{1,2} f_{\text{wall}}(\hat{\Delta}_l,\doublehat{\Delta}_2,\eta)\,,	
\end{equation}
where we defined the coefficients
\begin{equation}
	c_l^{1,2} =\mu^1_l \hat{\mu}^l_2\,,
\end{equation}
and the bulk-edge block
\begin{equation}
	f_{\text{wall}}(\hat{\Delta}_l,\doublehat{\Delta}_2,\eta)=(\eta+\eta^{-1})^{\doublehatexp{\Delta}_2}(2 x_{d-1})^{\hat{\Delta}_l}x_d^{\doublehatexp{\Delta}_2} \left( D[x_{d-1},\partial_{\vec{x}},\partial_{x_d}](2 x_d)^{-\hat{\Delta}_l + \doublehatexp{\Delta}_2}(2\hat{r}^2)^{-\doublehatexp{\Delta}_2}\right)|_{\vec{x}=0} \,,
\end{equation}
and we once more emphasized that we set $\vec{x}=0$ after applying the BOE.
Again, using the explicit expression for the Differential operator $D$, we get
\begin{equation}
\label{eq:bulkedgeblock}
f_{\text{wall}}(\hat{\Delta}_l,\doublehat{\Delta}_2,\eta) = \eta^{\hat{\Delta}_l-\doublehatexp{\Delta}_2} ~ _2F_1\left( \frac{\hat{\Delta}_l-\doublehat{\Delta}_2}{2},\frac{1+\hat{\Delta}_l-\doublehat{\Delta}_2}{2};\frac{3}{2}-\frac{d}{2}+\hat{\Delta}_l,-\eta^2\right)\,, 
\end{equation}
which clearly reduces to the one point block upon setting $\doublehat{\Delta}_2=0$, and is consistent with the OPE limit $\eta\to0$, as is easily checked by taking the leading term in eq. (\ref{eq:bulkedgefirstexp}).

Once again, we can also write down a Casimir equation that defines the block, and obtain it by imposing the OPE limit. We again write this in embedding space
\begin{equation}
\hat{L}^2\left( \frac{g_{\hat{\Delta}_l}\left(\frac{P\cdot V_1}{P\cdot V_2},\theta\right)}{(P\cdot V_1)^{\Delta_1-\doublehatexp{\Delta}_2}(-2 P_1 \cdot P_2)^{\doublehatexp{\Delta}_2}}\right)=c_{\hat{\Delta}_l,0}  \frac{g_{\hat{\Delta}_l}\left(\frac{P\cdot V_1}{P\cdot V_2},\theta\right)}{(P\cdot V_1)^{\Delta_1-\doublehatexp{\Delta}_2}(-2 P_1 \cdot P_2)^{\doublehatexp{\Delta}_2}}\,.
\end{equation}
Extracting the necessary prefactors, we derive an ODE for the function $g(\eta)$
\begin{equation}
\eta^2  \left(\eta ^2+1\right) g''(\eta )+\eta\left(2 \left(\eta ^2+1+\doublehat{\Delta}_2\right)-d\right) g'(\eta
	)+(\doublehat{\Delta}_2-\hat{\Delta}_l ) (\doublehat{\Delta}_2+\hat{\Delta}_l+1-d)g(\eta )=0\,.	
\end{equation}
Upon imposing the boundary condition $g(\eta)\sim \eta^{\hat{\Delta}_l - \doublehatexp{\Delta}_2}$ as $\eta$ approaches zero, we recover the block obtained in equation (\ref{eq:bulkedgeblock}).

As before, the ramp channel is obtained with the replacements
\begin{equation}
	x_{d-1} \to s_\perp \,,\, x_d \to s_\parallel \,,\, \eta \to \zeta\,.
\end{equation}
This leads to the bootstrap equation for the bulk-edge two point function
\begin{equation}
\label{eq:edgebulkboot}
	\sum_l c_l f_{\text{wall}}(\hat{\Delta}_l,\doublehat{\Delta}_2,\eta)= \left( \frac{\eta}{\sin \theta-\eta \cos \theta}\right) ^{\Delta_1-\doublehatexp{\Delta}_2}
	\sum_m c'_m f_{\text{ramp}}(\hat{\Delta}_m,\doublehat{\Delta}_2,\zeta)\,.
\end{equation}
It is clear that we recover the 1-pt bootstrap equation when taking $\doublehat{\Delta}_2=0$.
\subsection{Solutions with trivial boundaries}
We can begin checking the consistency of equation (\ref{eq:edgebulkboot}) by looking for solutions where the boundaries don't contain independent dynamics, which amounts to considering correlation functions obtained with one or even no boundaries. This corresponds to expanding a correlator in terms of our SO($d-1,1$) wedge blocks which in this case is a subgroup of the full symmetry.
Let us first take a 2-pt function in a homogeneous CFT
\begin{equation}
	\langle \mathcal{O}_1(\vec{0},x_{d-1},x_d) \mathcal{O}_2(0)\rangle = \frac{1}{r^{2\Delta_1}}\,.
\end{equation} 
Since in CFT two point functions are orthogonal, this means we set $\doublehat{\Delta}_2=\Delta_1$ in the prefactor of equation (\ref{eq:bulkedgeexp}). We must then have
\begin{equation}
	\sum_n c_n f_{\text{wall}}(\hat{\Delta}_l,\Delta_1,\eta)=	\sum_m c'_m f_{\text{ramp}}(\hat{\Delta}_r,\Delta_1,\zeta)=1\,.
\end{equation}
This is easily solved with a single block in each channel, by exchanging the operator $\hat{\Delta}= \Delta_1$. This is because of the truncation of the Hypergeometric series in the block
\begin{equation}
	f_{\text{wall}}(\Delta_1,\Delta_1,\eta)=f_{\text{ramp}}(\Delta_1,\Delta_1,\zeta)=1\,,
\end{equation}
where we also set $c_{\Delta_1}=c'_{\Delta_1}=1$.
We can also consider the slightly less trivial example of a single boundary at $x_{d-1}=0$. In this case we have a usual bulk-boundary 2-pt function of a BCFT which is fixed by kinematics to be
\begin{equation}
	\langle \mathcal{O}_1(\vec{0},x_{d-1},x_d) \hat{\mathcal{O}}_2(0)\rangle = \frac{1}{(2 x_{d-1})^{\Delta_1-\hat{\Delta}_2}r^{2\hat{\Delta}_2}}\,,
\end{equation}
where we set the bulk-boundary OPE coefficient to 1. In the wall channel we once again exchange only one operator $\hat{\Delta}_l=\hat{\Delta}_2$. However, we now have a non-trivial ratio of prefactors, and the crossing equation becomes
\begin{equation}
	1=\left( \frac{\eta}{\sin \theta-\eta \cos \theta}\right) ^{\Delta_1-\hat{\Delta}_2}
	\sum_m c'_m f_{\text{ramp}}(\hat{\Delta}_m,\hat{\Delta}_2,\zeta)\,,
\end{equation}
which is of course a generalization of the case studied in section \ref{sec:oneptsinglebdry}. Let us again, for simplicity, take $\theta=\pi/2$ and therefore expand around a virtual boundary at $x_d=0$. As in the one point function case, by expanding around $\eta \to \infty$ we find that we need an infinite tower of operators of the form $\hat{\Delta}_m = \Delta_1+2m$. The coefficients then read
\begin{equation}
	c'_m =\frac{4^{-m} \Gamma \left(\Delta _1+m+\frac{1}{2}-\frac{d}{2}\right)
		(\Delta_1-\hat{\Delta}_2)_{2 m}}{m! \Gamma \left(\Delta_1+2
	m+\frac{1}{2}-\frac{d}{2}\right)}\,,
\end{equation}
which clearly recover the one point function case upon setting $\hat{\Delta}_2=0$.
\subsection{Free bulk field}
We now return to solutions with non-trivial physics on both channels. Once again, it is a remarkable simplification to study the boundary and edge dynamics of a free bulk field $\phi$ which has dimension $\Delta_d= \frac{d}{2}-1$. Its correlation functions are defined by the free Schwinger-Dyson equations
\begin{equation}
	\square \langle \phi(\vec{x},x_{d-1},x_d) \dots \rangle = 0\,\,
\end{equation}
which holds at separated points. This will provide a nice check for the results obtained by solving the bootstrap equation.
It turns out that to solve crossing, the same boundary blocks $\hat{\Delta}= \frac{d}{2}-1 \,, \frac{d}{2}$ are enough even for generic $\doublehat{\Delta}_2$. Once again, the blocks dramatically simplify, and the crossing equation simply reads
\begin{align}
\label{eq:freebulkedgecrossing}
	&\frac{ c_{\hat{\phi}}  \cos \left(\left(\Delta_d-\doublehat{\Delta}_2\right)\phi  \right)+
		c_{\partial_\perp \hat{\phi}} \left(\Delta_d-\doublehat{\Delta}_2\right)^{-1} \sin \left(\left(\Delta_d-\doublehat{\Delta}_2\right)\phi  \right)}{\sin(\phi)^{\doublehatexp{\Delta}_2-\Delta_d}}=\nonumber\\
	&\frac{ c'_{\hat{\phi}}  \cos \left(\left(\Delta_d-\doublehat{\Delta}_2\right) (\theta
		-\phi )\right)+c'_{\partial_\perp \hat{\phi}} \left(\Delta_d-\doublehat{\Delta}_2\right)^{-1} \sin \left(\left(\Delta_d-\doublehat{\Delta}_2\right) (\theta
		-\phi )\right)}{\sin(\phi)^{\doublehatexp{\Delta}_2-\Delta_d} }\,.
	\end{align}
Amusingly, the solution to this crossing equation is trivial, as it is equivalent to the elementary trigonometric identities for the sum and difference of angles. 
We find
\begin{align}
\label{eq:freebulkedgesol}
c'_{\hat{\phi}} &= \cos\left(\left(\Delta_d-\doublehat{\Delta}_2\right)\theta\right)c_{\hat{\phi}}+\left(\Delta_d-\doublehat{\Delta}_2\right)^{-1}  \sin\left(\left(\Delta_d-\doublehat{\Delta}_2\right)\theta\right)c_{\partial_\perp \hat{\phi}}
\nonumber\\
c'_{\partial_\perp \hat{\phi}}&= \left(\Delta_d-\doublehat{\Delta}_2\right)\sin\left(\left(\Delta_d-\doublehat{\Delta}_2\right)\theta\right) c_{\hat{\phi}} - \cos\left(\left(\Delta_d-\doublehat{\Delta}_2\right)\theta\right) c_{\partial_\perp \hat{\phi}}\,.
\end{align}
As in the one point function case, these solutions can generically correspond to non-trivial boundary conditions, as we need a linear combination of both blocks to solve crossing.
However, we can now look for Dirichlet-Dirichlet solutions where $c_{\hat{\phi}}=c'_{\hat{\phi}}=0$. This solution is the starting point for the perturbative analysis of Cardy in $4-\epsilon$ dimensions \cite{Cardy_1983}.
The edge dimension gives us enough room to impose Dirichlet boundary conditions for arbitrary $\theta$. This leads to the following constraint on $\doublehat{\Delta}_2$
\begin{equation}
	\doublehat{\Delta}_{2} = \frac{d}{2}-1 +n \frac{\pi}{\theta}\,, 
\end{equation}
with $n$ an arbitrary integer. Additionally the expansion coefficients are constrained to satisfy $c'_{\partial_\perp \hat{\phi}}=c_{\partial_\perp \hat{\phi}}$. For Dirichlet boundary conditions in the normal BCFT setup where $\theta=\pi$ , the boundary operator should just be interpreted as $\partial_\perp \hat{\phi}$, meaning $\doublehat{\Delta}_2(\theta=\pi)=\frac{d}{2}$. We then conclude that\footnote{Note that the operators with negative $n$ are non-unitary, as their dimension can be made arbitrarily negative by making $\theta$ small. The $n=1$ operator is the most relevant and therefore determines the critical exponents in gaussian theories.}:
\begin{equation}
	\doublehat{\Delta}_{DD} = \frac{d}{2}-1 +\frac{\pi}{\theta}\,,
\end{equation}
as obtained by Cardy in \cite{Cardy_1983}. Remarkably, this captures a non-trivial anomalous dimension, although we are studying a free theory with free boundary conditions. The final correlator is quite simple:
\begin{equation}
	\langle \mathcal{O}_1(\vec{0},x_{d-1},x_d) \doublehat{\mathcal{O}}_2(0)\rangle_{DD} = \frac{\sin\left( \frac{\pi \phi}{\theta}\right) }{r^{d-2+\frac{\pi}{\theta}}}\,,
\end{equation}
where we set the overall free coefficient to 1.

It is not hard to solve the crossing equations for other free boundary conditions. For example setting $c'_{\partial_\perp \hat{\phi}}=c_{\partial_\perp \hat{\phi}}=0$, which is Neumann-Neumann gives
\begin{equation}
\doublehat{\Delta}_{NN} = \frac{d}{2}-1 +\frac{2\pi}{\theta}\,.
\end{equation}
We can also consider Dirichlet-Neumann boundary conditions and obtain
\begin{equation}
\doublehat{\Delta}_{ND} = \frac{d}{2}-1 +\frac{\pi}{2\theta}\,.
\end{equation}
We can also reproduce the general solution for an arbitrary combination of Dirichlet and Neumann blocks, through the use of the equations of motion, as mentioned above.
We have
\begin{equation}
	\square \langle \phi(\vec{x},x_{d-1},x_d) \doublehat{\mathcal{O}}(0) \rangle =0\,,
\end{equation}
we will eventually set $\vec{x}=0$ but only after acting with the laplacian. Specifying the kinematical structure of the correlator leads to
\begin{equation}
	\left( \frac{\partial^2}{\partial x_{d-1}^2}+\frac{\partial^2}{\partial x_{d}^2}+\frac{\partial^2}{\partial \vec{x}^2}\right) \frac{g\left( \frac{x_{d-1}}{x_d}\right) }{(2 x_{d-1})^{\Delta_1-\doublehatexp{\Delta}_2}r^{2\doublehatexp{\Delta}_2}}=0\,,
\end{equation}
which leads to the ODE
\begin{equation}
4 \eta^2  \left(\eta
^2+1\right) g''(\eta )+4 \eta\left(2 \left(\doublehat{\Delta}_2+\eta ^2+1\right)-d\right) g'(\eta )+\left(d-2 \doublehat{\Delta}_2\right) \left(d-2 \left(\doublehat{\Delta}
_2+1\right)\right) g(\eta)=0\,.	
\end{equation}
The two independent solutions to this equation are once again the Neumann and Dirichlet block, and we can of course take the most general solution to be a combination of both.

\subsubsection{Comments on the order $\epsilon$ bootstrap}
These simple solutions are interesting as they can be a starting point for perturbative expansions. In particular, Cardy studied the $\epsilon$ expansion to first order with $DD$ boundary conditions \cite{Cardy_1983}. Let us briefly comment on how this fits into our framework. First, we recall that in the BCFT 2-pt function bootstrap, the order $\epsilon$ correlator can still be obtained with a finite sum of blocks \cite{Liendo:2012hy}. Of our particular interest is the boundary channel expansion. In this channel, for Dirichlet boundary conditions, we still only exchange the operator $\partial_\perp \hat{\phi}$, although it acquires an order $\epsilon$  anomalous dimension, and there is an order $\epsilon$ correction to the expansion coefficient. This may lead one to believe that we can solve our crossing equation around Dirichlet boundary conditions at order $\epsilon$ by still exchanging only $\partial_\perp \hat{\phi}$. A simple ansatz to first order in $\epsilon$, allowing only for order $\epsilon$ corrections to the CFT data of the order zero solution fails to give a non-trivial result. After a moment's thought, one remembers the existence of an infinite tower of boundary operators of dimension $2n+2$ contributing at order $\epsilon^2$ to the BCFT bootstrap. Since the expansion coefficient in this case is the square of the bulk-boundary OPE coefficient, this means that the bulk-boundary coefficient is of order $\epsilon$. In the boundary case, the square increases the order in $\epsilon$ from one to two, leading to the fact that only operators that already appeared at order zero can appear at first order \cite{Liendo:2012hy,Bissi:2018mcq}. In the wedge setup such a simplification does not happen. This is because our expansion coefficient is a product $\mu^1_l \hat{\mu}^l_2$, which contains one bulk to boundary and one boundary to edge coefficient. As we argued, the bulk to boundary coefficients for the Dirichlet operators are of order $\epsilon$, but we generally allow the boundary to edge coefficients to be of order one, meaning our correlator should contain infinitely many blocks already at order $\epsilon$. The diagrammatic calculation of Cardy seems to support this possibility, as is visible by the infinite number of contributions that must be taken into account in the two point correlator. We note, however, that Cardy was able to isolate the relevant logarithmic singularity and obtain the edge anomalous dimension $\doublehat{\gamma}_2$, which we quote here for the O($N$) model \cite{Cardy_1983}
\begin{equation}
	\doublehat{\gamma}_2=- \frac{N+2}{2(N+8)} \frac{(5\pi^2/\theta^2+1)}{6\pi/\theta}\,,
\end{equation}
notably, this expression reproduces the anomalous dimension of $\partial_\perp \hat{\phi}$ for $\theta=\pi$. To reproduce this result, we need techniques to handle the infinite sums of blocks. Such techniques were used in the BCFT bootstrap to obtain order $\epsilon^2$ results \cite{Bissi:2018mcq} and it should be possible to adapt them to the order $\epsilon$ problem in our setup. We leave this exploration for future work.
\subsection{Generalized free field solution}
Upon a careful observation of the crossing equation for a free bulk field, eq. (\ref{eq:freebulkedgecrossing}), and its solution eq. (\ref{eq:freebulkedgesol}), we notice that the fact that the dimension of the external bulk field was the free field dimension $\Delta_d=\frac{d}{2}-1$ isn't particularly important. In fact, performing the  formal replacement $\Delta_d\to \Delta_1 $ we find a generalized free field solution:
\begin{equation}
	\langle \mathcal{O}_1(\vec{0},x_{d-1},x_d) \doublehat{\mathcal{O}}_2(0)\rangle_{\text{GFF}} = \frac{ c_{\hat{\phi}} 
			\cos \left(\left(\Delta _1-\doublehat{\Delta}_2\right) \phi \right)+c_{\partial_\perp \hat{\phi}} (\Delta_1-\doublehat{\Delta}_2)^{-1} \sin
			\left(\left(\Delta _1-\doublehat{\Delta}_2\right) \phi \right)}{\sin(\phi )^{\doublehatexp{\Delta}_2-\Delta _1}(2 x_{d-1})^{\Delta_1-\doublehatexp{\Delta}_2}r^{2\doublehatexp{\Delta}_2}}\,,
\end{equation}
which is crossing symmetric, and remarkably simple. However the simplification happens only at the level of the correlation function, since the individual blocks only simplify for dimensions that are integer separated from a free field. In particular, expanding the invariant part of the correlator at small $\eta$ we find the behaviour $g(\eta)\sim \eta^{\Delta_1-\doublehatexp{\Delta}_2}(1+\eta+ O(\eta^2))$. We then find that the decomposition of this correlator in wall channel blocks corresponds to an infinite tower of operators of dimensions $\hat{\Delta}_n$ such that
\begin{equation}
	\hat{\Delta}_n = \Delta_1+n \,, n \in \mathbb{Z}_{\geq0}\,.
\end{equation}
Without loss of generality, we can set the overall coefficients $c_{\hat{\phi}}= c_{\partial_\perp \hat{\phi}}=1$, and find the coefficients $c_n$ for each of the operators exchanged in the boundary.
We obtain
\begin{align}
	c_n &= \frac{(-1)^{n/2} \left(\Delta _1-\doublehat{\Delta} _2\right)_n
		\left(-\frac{d}{2}+\Delta _1+\frac{1}{2}\right)_{\frac{n}{2}} \left(-\frac{d}{2}+\Delta
		_1+1\right)_{\frac{n}{2}}}{2^n n! \left(\frac{1}{4} \left(-d+2 \Delta
		_1+1\right)\right)_{\frac{n}{2}} \left(\frac{1}{4} \left(-d+2 \Delta
		_1+3\right)\right)_{\frac{n}{2}}}\,, \quad n \,\text{even}\\
		c_n &=\frac{(-1)^{(n-1)/2} \left(\Delta _1-\doublehat{\Delta} _2+1\right)_{n-1} \left(-\frac{d}{2}+\Delta
			_1+1\right)_{\frac{n-1}{2}} \left(-\frac{d}{2}+\Delta
			_1+\frac{3}{2}\right)_{\frac{n-1}{2}}}{ 2^{n-1}n! \left(\frac{1}{4} \left(-d+2 \Delta
			_1+3\right)\right)_{\frac{n-1}{2}} \left(\frac{1}{4} \left(-d+2 \Delta
			_1+5\right)\right)_{\frac{n-1}{2}}}
		\,, \quad n \,\text{odd}\,.
\end{align}
On the ramp/floor channel, we again have infinitely many operators of the form $\hat{\Delta}_m= \Delta_1 +m$, with some $\theta$ dependent coefficients $c'_m$. This is the simplest solution with infinitely many operators on both channels. We also note that we can obtain a GFF type one point function by setting $\doublehat{\Delta}_2=0$. 
\section{Conclusions}
\label{sec:Conclusions}
In this work, we developed the necessary machinery to start a bootstrap program for correlators of a CFT in a wedge configuration with angle $\theta$ between the intersecting boundaries. We studied the kinematics of bulk, boundary and edge correlation functions, emphasizing the bulk one point function, and the bulk-edge two point function, which are the simplest non-trivial correlators, depending on a cross-ratio $\eta=\tan(\phi)$ and the parameter $\theta$. 

We developed a conformal block expansion for these correlation functions, taking advantage of the convergence of the boundary operator expansion. We obtained explicit expressions for the blocks using the BOE and the Casimir equation. Imposing the equality of the  two boundary expansions lead us to a one parameter family of non-perturbative crossing equations, analogous to many others in the CFT literature. We analytically solved these equations in simple cases, namely for fictitious boundaries, for generalized free fields, and for a free bulk field. 

The case of a free bulk field is of particular interest for applications, since it provides a starting point for perturbative expansions, for example the $\epsilon$ expansion. We were able to obtain the leading dimension for the edge operator under free boundary conditions, reproducing and extending results by Cardy \cite{Cardy_1983}. We also obtained the general solution where the boundary theory contains an arbitrary linear combination of the Neumann and Dirichlet operators.

There several open directions to build open the basic framework we developed. The most obvious one is the analysis of the analytic structure of the blocks and study of discontinuities of the crossing equation, or more general dispersive techniques, which have the potential to address the infinite sums of blocks that appear in the $\epsilon$ expansion at first order. The techniques developed by \cite{Bissi:2018mcq,Dey:2020jlc} have the potential to be transported to this context. The anomalous dimension of the edge operator obtained by Cardy seems like the perfect benchmark to test the full potential of our setup. 

Another avenue is to study the crossing equation non-perturbatively, through the use of numerical techniques such as linear or semi-definite programming \cite{Rattazzi:2008pe,Kos_2014,Simmons-Duffin:2015qma}. A first obstacle to this is that we do not have manifest positivity of the expansion coefficients in either channel. One could of course take this positivity as an input and study the numerical bounds with the understanding that their applicability is limited. An obvious target would be the 3d Ising model, or even the $\epsilon$ expansion, since the dependence on space-time dimension of the blocks is very mild, as in the BCFT bootstrap \cite{Liendo:2012hy}. An alternative that bypasses the sign problem of the coefficients is to use a Gliozzi type method of determinants \cite{Gliozzi:2015qsa,Gliozzi:2013ysa}, although this technique has other limitations, since one cannot use it to obtain rigorous error bars.

There is also a potential relation to holographic physics \cite{Maldacena:1997re,Witten:1998qj,Gubser:1998bc}. There are several similar (but different) holographic setups where a wedge plays a role. We find of note, the wedge holography between AdS$_{d+1}$ and CFT$_{d-1}$ of \cite{Akal:2020wfl}, the interface-type holography studied in \cite{Bachas:2020yxv,Bachas:2021fqo} and others in the entanglement entropy literature \cite{Bianchi:2015liz,Bianchi:2016xvf,Geng:2020fxl}. For a more direct relation it would be interesting to construct a holographic setup dual to the wedge configuration. This would imply considering a system with a set of AdS$_{d+1}$, AdS$_{d}$, AdS$_{d-1}$ spaces and the dual CFT$_d$, CFT$_{d-1}$ and CFT$_{d-2}$ that we have considered. The language and formalism of \cite{Rastelli:2017ecj}, where several Witten diagrams dual to BCFT/ICFT were computed, can potentially be generalized to allow for one more co-dimension \cite{Goncalves:2018fwx}, embedding our setup into their calculations. This also suggests that Mellin amplitudes could be a useful tool to study our wedge correlators, at least if they are of holographic nature.

Finally, we mention that systems of several boundaries and defects are very common in the literature of supersymmetric, and in particular superconformal field theories. Notably, in the context of the SCFT-chiral algebra correspondence \cite{Beem:2013sza} there have been recent studies of setups with intersecting defects \cite{Gomis:2016ljm,Pan:2016fbl}. It would be interesting to see if our program can be generalized to intersecting defects of arbitrary co-dimension, and if the bootstrap approach can give further insight into the dynamics of these systems.

\section*{Acknowledgements}
\label{sec:ack}
We are happy to thank Miguel Costa, Vasco Gonçalves, Tobias Hansen, Edoardo Lauria, Marco Meineri and Sourav Sarkar for helpful discussions and Miguel Costa, Vasco Gonçalves, Tobias Hansen, Edoardo Lauria and Sourav Sarkar for comments on a draft. We are particularly indebted to Edoardo Lauria for inspirational discussions in the intermediate stages of this work and to Sourav Sarkar for a thorough reading of a draft.
This research received funding from the Simons Foundation grants 488637  (Simons collaboration on the Non-perturbative bootstrap).
Centro de F\'\i sica do Porto is partially funded by Funda\c c\~ao para a Ci\^encia e a Tecnologia (FCT) under the grant
UID-04650-FCUP.
AA is funded by FCT under the IDPASC doctoral program with the fellowship  PD/BD/\allowbreak 135436/2017.
 
\bibliographystyle{JHEP}
\bibliography{EdgeCFT.bib}

\end{document}